\documentclass[%
 reprint,
 amsmath,amssymb,
floatfix,
]{revtex4-2}
\usepackage[T1]{fontenc}
\usepackage[utf8]{inputenc}
\usepackage{latexsym}
\DeclareMathSizes {11}{9.4}{8}{8}
\usepackage{epsfig}
\usepackage{subfig}
\usepackage[justification=centering]{caption}
\usepackage{color}
\usepackage{float}
\usepackage{natbib}
\usepackage{graphicx}
\usepackage{array}
\usepackage{microtype }
\usepackage{epstopdf,amsmath}
\usepackage{perpage} 
\usepackage{footnote}
\usepackage{multirow}
\newcolumntype{C}[1]{>{\centering\arraybackslash}m{\dimexpr#1-2\tabcolsep\relax}}
\makeatletter
\newcommand*{\rom}[1]{\expandafter\@slowromancap\romannumeral #1@}
\makeatother
\makeatletter
\def\blfootnote{\xdef\@thefnmark{}\@footnotetext}
\makeatother
\usepackage{dcolumn}
\usepackage{subfig}
\usepackage{bm}

\begin{document}
\preprint{APS/123-QED}

\title{ A note on inflation in dRGT massive gravity}
\author{{B. Afshar$^1$}}
\email{afshar.behnoush@gmail.com} 
\author{N. Riazi$^1$} %
 \email{n$\_$riazi@sbu.ac.ir} %
 \author{H. Moradpour$^2$}
  \email{hn.moradpour@maragheh.ac.ir}
\affiliation{%
 $^1$ Department of Physics, Shahid Beheshti University, Tehran 19839, Iran\\
$^2$ Research Institute for Astronomy and Astrophysics of Maragha
(RIAAM), University of Maragheh, P.O. Box 55136-553, Maragheh,
Iran
}%

\date{\today}

\begin{abstract}
Although the dRGT massive gravity successfully explains the late-time
cosmic acceleration, it cannot justify inflation. On the other
hand, and in the frameworks of General Relativity and modified
gravity, the interests and attempts to describe dark energy and
inflation by using Lagranginas, which may have pole, have recently been enhanced. Subsequently, we are going to show that this kind of Lagrangian may justify inflation in the framework of dRGT massive gravity. The study is done focusing on the power and exponential potentials,
and the results show a plausible consistency with the Planck 2018 data and its combination with BK18 and BAO.
\end{abstract}

\maketitle
\section{Introduction}

Despite its enormous successes, General Relativity (GR) cannot be
considered as the final theory of gravity, and needs to be
modified \cite{1}. Two of the most important motivations for
modifying GR are the explanation of inflation, and the late-time
cosmic acceleration \cite{2}. Massive gravity theory is an
approach to modify GR which includes a graviton of small mass
\cite{3}. In 1939, Fierz and Pauli (FP) \cite{4} proposed
massive gravity theory for the first time. By adding the
fine-tuned mass term to the linearized Einstein-Hilbert action,
they developed a theory that correctly leads to 5 degrees of
freedom for the massive spin-2 particle \cite{1,5}. In 1970,
van Dam, Veltman \cite{6}, and Zakharov (vDVZ) \cite{7} discovered
that, in the massless limit, the FP theory does not render GR
\cite{5}. In other words, the principle of conformity is violated
leading to a deviation in the gravitational lensing prediction of
theory about 25\% compared to that of GR \cite{8,9}. This
discontinuity is known as the vDVZ discontinuity \cite{8}.

Focusing on the nonlinear extensions of the FP theory, Vainshtein
\cite{10} found out a distance scale, called the Vainshtein
radius
${{r}_{V}}={{(\frac{M}{M_{Pl}^{4}m_{g}^{4}})}^{\frac{1}{5}}}$,
where $M_{Pl}$ is the Planck mass, and $m_g$ denotes the graviton
mass) \cite{11,12,13}. For $r<{{r}_{V}}$, the nonlinear effects
become highly important \cite{11,12,13} meaning that the outcomes
differ from those predicted by FP. As the graviton mass becomes
less, the ${r}_{V}$ is boosted and at the massless limit,
${r}_{V}$ tends to infinity. Consequently, the results of the FP
theory are no longer valid \cite{11,12,13}. So, it seems possible
to rectify the vDVZ discontinuity by considering the nonlinear
effects \cite{11,12,13}, an approach called the Vainshtein
mechanism \cite{13}.

According to the Vainshtein's idea, Boulware and Deser (BD)
\cite{14} find that all nonlinear extensions of the FP theory have
an additional degree of freedom (a ghost-like scalar mode), known
as the BD ghost \cite{15,16}. The discovery of the late-time
cosmic acceleration in 1998, and the unknown nature of
cosmological constant led scientists to investigate modified
gravity theories, such as massive gravity theory, more precisely.
It is also realized that a non-renormalizable theory, such as the
FP theory and even one with apparent instabilities, can be
understood as an effective field theory, valid only at energies
below an ultraviolet cutoff scale \cite{17}.

In 2003, Arkani-Hamed, Georgi, and Schwartz \cite{18} could
return the gauge invariance to massive gravity theory by employing
the St$\ddot{\text{u}}$ckelberg trick leading to develop an
effective field theory for massive gravity. The
St$\ddot{\text{u}}$ckelberg trick also helps us solve vDVZ
discontinuity. In the massless limit, the effects of the strong
decoupling neutralize the ghost-like scalar mode, and consequently
the results of GR are reproduced. Moreover, in certain nonlinear
extensions of FP, the cutoff scale ${\Lambda_{5}}={{(
{{M}_{Pl}}m_{g}^{4})}^{{}^{1}/{}_{5}}}$ of FP can also be
raised to ${{\text{}\!\!\Lambda\!\!\text{ }}_{3}}={{(
{{M}_{Pl}}m_{g}^{2})}^{{}^{1}/{}_{3}}}$. De Rham, and
Gabadadze \cite{19} determine the coefficients of higher-order
interaction terms in a way that the obtained action, as a
generalization of FP, does not include the BD ghost in the
decoupling limit. One year later, in 2011, De Rham, Gabadadze,
and Tolley \cite{20} could propose a covariant nonlinear theory of
massive gravity that correctly describes the massive spin-2
field; the birth of dRGT theory. The dRGT theory is indeed
ghost-free in the decoupling limit to all orders.

Although dRGT theory does not have flat, and closed FRW solutions,
the open FRW solution yields an effective cosmological constant
proportional to the graviton mass (${{m}_{g}}$) \cite{21}. In the
report of Ligo-Virgo Collaboration about the discovery of the
first gravitational wave, the constraint ${{m}_{g}}<1.2\times
{{10}^{-22}}eV$ is determined \cite{22}. If ${{m}_{g}}\sim
{{H}_{0}}\sim {{10}^{-33}}eV$, then the effective cosmological
constant can explain the late-time cosmic acceleration \cite{23}.
Since the graviton mass cannot be of the order of the Hubble
constant during the inflation, dRGT theory is not capable to
produce enough number of e-foldings \cite{24}.

Recently, a new type of Lagrangian, which may include pole and thus called pole Lagrangian, has attracted attempts to itself in order to model and describe the current and primary inflationary eras \cite{25,26,27,28}. This kind of Lagrangian even
supports evolving traversable wormholes satisfying energy
conditions in the framework of GR \cite{29}. Motivated by these attampets and also Ref.~\cite{24}, here, we use the pole Lagrangian to build some models for the primary inflationary era in the framework of dRGT theory.

The action of the scalar field $\sigma$ is written as

\begin{equation}
S_\sigma=\int{{d^4}x}\sqrt{-g}[-\frac{1}{2}f(\sigma){g^{\mu \nu }}{{\partial }_{\mu }}{{\sigma}^{{}}}
{{\partial }_{\nu }}\sigma -V(\sigma )],
\end{equation}

\noindent where $V(\sigma)$ denotes the potential and $f(\sigma)$ is a function of the scalar field $\sigma$ \cite{30}. The kinetic term can also be brought into canonical form

\begin{equation}
S_\varphi =\int{{d^4}x}\sqrt{-g}[-\frac{1}{2}{{g}^{\mu
\nu }}{{\partial }_{\mu }}{{\varphi }^{{}}}{{\partial }_{\nu
}}\varphi -V(\varphi)],
\end{equation}
 \noindent using the transformation
 \begin{equation}
d\varphi=\sqrt{f(\sigma)}d\sigma,
 \end{equation}	
\noindent leading to \cite{31}
\begin{equation}
\varphi(\sigma)=\int_{\sigma_0}^{\sigma}{d{\sigma}} \sqrt{f({\sigma})},
\end{equation}

\noindent in which $\sigma_0$ denotes the desired point, which is considered as ${\sigma_0}=0$, here \cite{31}. If $f(\sigma)$ is singular at some points, i.e., then it  possesses pole, it is impossible to integrate from $f(\sigma)$ \cite{31}. Assuming pole is located in $\sigma=\sigma_p$, if $\sigma_0 < \sigma_p$, then $\sigma$ lies within the branch $(-\infty,{\sigma_p})$. Also, when $\sigma_0 > \sigma_p$, it lies within the branch $({\sigma_p},\infty)$ \cite{31}. To provide a general outlook on the possibility of describing inflation
in the dRGT theory, we first derive the corresponding formulation
for the general potential $V(\sigma)$ and the general function $f(\sigma)$. Then, we concentrate on the power and exponential potentials, as well as the particular function $f(\sigma)={A}{{|\sigma|}^{-p}}$ which attended in Refs.~\cite{25,26,27,28,29,31}.

The paper is organized as follows. The dRGT theory and corresponding open FRW solutions are briefly reviewed in Sec. (II) and (III), respectively. In the continuation of the third section, the number of e-foldings $N_e$, scalar spectral
	$n_s$, and tensor-to-scalar ratio $r$ are also derived. One of our main goals in Sec. (IV) is to show the power of models in producing the appropriate number of e-foldings to solve the flatness problem, as well as the consistency of models with the Planck 2018 data and its combination with BK18 and BAO is investigated through employing $r(n_s)$. Special solutions that lead to the invalidation of models are also discussed. A summary is finally presented in the last section. It is assumed that ${M_{Pl}}=1$.

\section{The $\textbf{d}$RGT Theory}

The dRGT theory is described by the physical metric ${{g}_{\mu
\nu}}$, and the non-dynamical (or fiducial) metric ${{f}_{\mu \nu
}}$ so that

\begin{equation}
{{g}_{\mu \nu }}={{f}_{\mu \nu }}+{{H}_{\mu \nu }}.
\end{equation}

\noindent Here, ${{H}_{\mu \nu }}$ denotes the covariantization of
the metric perturbation (${{h}_{\mu \nu }}\equiv {{g}_{\mu \nu
}}-{{\eta }_{\mu \nu }}$), and the fiducial metric is defined as

\begin{equation}
{{f}_{\mu \nu }}={{\tilde{f}}_{ab}}{{\partial }_{\mu }}{{\phi
}^{a}}{{\partial }_{\nu }}{{\phi }^{b}},
\end{equation}

\noindent where ${{\tilde{f}}_{\mu \nu }}$ is the reference
metric, and ${{\phi }^{a}}(a=0,1,2,3)$ are the four fields
introduced to restore general covariance. For a Minkowskian
reference metric, ${{\phi }^{a}}$ form a 4-vector, implying that
the dRGT theory carries the global Poincare symmetry \cite{32}.\\
\noindent The gravitational action consists of the Einstein-Hilbert term,
and the mass term

\begin{equation}
    \begin{aligned}
& {{S}_{g}}={{S}_{EH}}+{{S}_{mass}}, \\
& {{S}_{g}}=\int{{{d}^{4}}x}\sqrt{-g}[\frac{R}{2}+m_{g}^{2} \mathcal{U}\left( {{g}_{\mu \nu }},{{H}_{\mu \nu }} \right)], \\
\end{aligned}
\end{equation}

\noindent in which $R=g_{{}}^{\mu \nu }R_{\mu \nu }^{{}}$ is
the Ricci scalar, and $\mathcal{U}$ is the potential without
derivatives of terms originated from the interaction between
${{H}_{\mu \nu }}$ and ${{g}_{\mu \nu }}$ \cite{23}. To prevent
the appearance of the BD ghost, $\mathcal{U}$ is constructed as

\begin{equation}
\mathcal{U}\left( {{g}_{\mu \nu }},{{H}_{\mu \nu }}
\right)=\sum\limits_{n=0}^{4}{{{\alpha
}_{n}}}{{\mathcal{L}}_{\text{n}}}.
\end{equation}

\noindent In this formula, ${{\alpha }_{n}}$ are the dimensionless
parameters, and

\begingroup\makeatletter\def\f@size{9}\check@mathfonts
\begin{equation}
\begin{aligned}
& {{\mathcal{L}}_{0}}=1,~ \\ 
& {{\mathcal{L}}_{1}}=\left[ \mathcal{K} \right], \\ 
& {{\mathcal{L}}_{2}}=\frac{1}{2}~\left( {{\left[ \mathcal{K} \right]}^{2}}-\left[ {{\mathcal{K}}^{2}} \right] \right), \\ 
& {{\mathcal{L}}_{3}}=~\frac{1}{6}\left( {{\left[ \mathcal{K} \right]}^{3}}-3\left[ \mathcal{K} \right]\left[ {{\mathcal{K}}^{2}}\left] +2 \right[{{\mathcal{K}}^{3}} \right] \right), \\ 
& {{\mathcal{L}}_{4}}=\frac{1}{24}~\left( {{\left[ \mathcal{K} \right]}^{4}}-6{{\left[ \mathcal{K} \right]}^{2}}\left[ {{\mathcal{K}}^{2}} \right]+3{{\left[ {{\mathcal{K}}^{2}} \right]}^{2}}+8\left[ \mathcal{K} \right]\left[ {{\mathcal{K}}^{3}}\left] -6 \right[{{\mathcal{K}}^{4}} \right] \right), \\ 
\end{aligned}
\end{equation}
\endgroup

\noindent where $\mathcal{K}_{~\nu }^{\mu }$ is defined as

\begin{equation}
\mathcal{K}_{~\nu }^{\mu }=\delta _{_{{}}\nu }^{\mu
}-\sqrt{{{g}^{\mu \lambda }}{{f}_{\lambda \nu }}},
\end{equation}

\noindent and the square brackets denote trace operation i.e.,

\begin{equation}
\begin{aligned}
& \left[ \mathcal{K} \right]=\mathcal{K}_{~\mu }^{\mu },~~~~~~~~~~~~\left[ {{\mathcal{K}}^{2}} \right]=\mathcal{K}_{~\nu }^{\mu }\mathcal{K}_{~\mu }^{\nu },~~ \\
& \left[ {{\mathcal{K}}^{3}} \right]=\mathcal{K}_{~\nu }^{\mu }\mathcal{K}_{~\lambda }^{\nu }\mathcal{K}_{~\mu }^{\lambda },\text{ }\left[ {{\mathcal{K}}^{4}} \right]=\mathcal{K}_{~\nu }^{\mu }\mathcal{K}_{~\lambda }^{\nu }\mathcal{K}_{~\rho }^{\lambda }\mathcal{K}_{~\mu }^{\rho }.~ \\
\end{aligned}
\end{equation}

\noindent ${{g}_{\mu \nu }}$ raises, and lowers the indices, and
moreover, ${\mathcal{L}}_{0}$ corresponds to a cosmological
constant, ${\mathcal{L}}_{1}$ and ${\mathcal{L}}_{2}$ to a
tadpole, and the mass term, respectively. ${\mathcal{L}}_{3,4}$
also include higher-order interaction terms \cite{16}. Setting
${{\alpha }_{0}}={{\alpha }_{1}}=0$ Minkowskian spacetime becomes
a vacuum solution \cite{16}, and in addition, if ${{\alpha
}_{2}}=1$, then the FP theory, at a linearized level, is recovered
\cite{33}. ${{\alpha }_{3}}$, and ${{\alpha }_{4}}$ are the free
parameters of the dRGT theory. Accordingly, the gravitational action
(7) is finally rewritten as

\begin{equation}
\!{{S}_{g}}=\int{{{d}^{4}}x}\sqrt{-g}[\frac{R}{2}+m_{g}^{2}({{\mathcal{L}}_{2}}\text{
}\!\!~\!\!\text{ }+{{\text{ }\!\!\alpha\!\!\text{
}}_{3}}{{\mathcal{L}}_{3}}\text{ }\!\!~\!\!\text{ }+{{\text{
}\!\!\alpha\!\!\text{ }}_{4}}\text{ }\!\!~\!\!\text{
}{{\mathcal{L}}_{4}}\text{ }\!\!~\!\!\text{ })].
\end{equation}

\section{BACKGROUND COSMOLOGY}

The total action consists of three parts including $i$) the
gravitational action (12), $ii$) the Lagrangian (1), and
$iii$) the matter action $S_m$ so that

\begin{equation}
{{S}_{tot}}={{S}_{g}}+{{S}_{\sigma}}+{{S}_{m}},
\end{equation}

\noindent where $S_m$ corresponds to a perfect fluid, described by
the energy-momentum tensor

\begin{equation}
T_{~\nu }^{\mu }=diag(-{{\rho
}_{m}},{{p}_{m}},{{p}_{m}},{{p}_{m}}),
\end{equation}

\noindent in which ${{\rho }_{m}},$ and ${{p}_{m}}$ denote the
energy density, and the pressure of matter, respectively.

Now, assuming the Universe is homogeneous and isotropic, the
physical metric ${{g}_{\mu \nu }}$ is deemed as the Friedmann-
Robertson-Walker (FRW) metric

\begin{equation}
{{g}_{\mu \nu }}d{{x}^{\mu }}d{{x}^{\nu
}}=-N{{(t)}^{2}}{{(dt)}^{2}}+a{{(t)}^{2}}{{\Omega
}_{ij}}d{{x}^{i}}d{{x}^{j}}.
\end{equation}

\noindent Here, $N$ is the cosmological time, $a$ is the scale
factor, and ${{\Omega }_{ij}}$ denotes the metric of unit
$3$-sphere

\begin{equation}
    {{\Omega }_{ij}}={{\delta }_{ij}}+\frac{k}{1-k{{r}^{2}}}{{x}^{i}}{{x}^{j}},
\end{equation}

\noindent where $k$ is the spatial curvature, and
${{r}^{2}}={{x}^{2}}+{{y}^{2}}+{{z}^{2}}$. 
\noindent In this paper, the Minkowskian form of the reference
metric is chosen as

\begin{equation}
{{\tilde{f}}_{\mu \nu }}={{\eta }_{\mu \nu }}=diag(-1,1,1,1).
\end{equation}

As demonstrated in Ref \cite{16}, the flat FRW solution of the
dRGT theory is equal to $a=$const, inconsistent with a dynamic
Cosmos. Furthermore, the dRGT theory lacks closed FRW solutions,
since the fiducial Minkowski metric cannot be foliated by closed
slices \cite{21}. Therefore, we only focus on open FRW of negative
$k$ very close to zero, in accordance with the Planck
collaboration prescription \cite{34,35}. In order to carry the
symmetries of an open FRW metric by the fiducial metric, the
St$\ddot{\textmd{u}}$ckelberg fields should be chosen as

\begin{equation}
{{\phi }^{0}}=f(t)\sqrt{1-k{{r}^{2}}},\quad {{\phi
}^{i}}=\sqrt{-k}f(t){{x}^{i}},
\end{equation}

\noindent leading to

\begin{equation}
{{f}_{\mu \nu }}d{{x}^{\mu }}d{{x}^{\nu }}=-{{\dot{f}(t)}^{2}}{{(dt)}^{2}}-kf{{(t)}^{2}}{{\Omega }_{ij}}d{{x}^{i}}d{{x}^{j}},
\end{equation}

\noindent where $\dot{f}=\frac{df(t)}{dt}$. Without loss of
generality, we also assume $a>0,~N>0,~f\ge 0,~\dot{f}\ge 0$.

\subsection{Equations of motion}

Inserting the physical metric~(15), and the fiducial metric~(19)
in Eq.~(13), the total action is obtained as

\begin{align}\notag
S_{tot}&=\int d^{4}x\bigg[3\big[-\frac{a\dot{a}^{2}}{N}+kNa+m_{g}^{2}\big(NF(a,f) \\ \label{20}
&  -\dot{f}G(a,f)\big)\big]+a^{3}\big(\frac{1}{2N}f(\sigma)\dot{\sigma }^{2}-{N}V(\sigma )\big)\bigg]+S_m, 
\end{align}

\noindent in which

\begin{equation}
\begin{aligned}
&F(a,f)=a(a-\sqrt{-k}f)(2a-\sqrt{-k}f)+~~~~\\
& \frac{{{\alpha }_{3}}}{3}{{(a-\sqrt{-k}f)}^{2}}(4a-\sqrt{-k}f)+\frac{{{\alpha }_{4}}}{3}{{(a-\sqrt{-k}f)}^{3}},~~~~~~~~~ \\
\end{aligned}
\end{equation}

\noindent and

\begin{equation}
\begin{aligned}	
&G(a,f)={{a}^{2}}(a-\sqrt{-k}f)+\alpha _{3}^{{}}a{{(a-\sqrt{-k}f)}^{2}}+\\
&\frac{{{\alpha }_{4}}}{3}{{(a-\sqrt{-k}f)}^{3}}.
\end{aligned}
\end{equation}
\\
\noindent Now, variation of the total action~(20) with respect to
the field variables $f,~N,~a$, and$~\sigma$ renders

\begin{equation}
{{\delta}_{f}}{{S}_{tot}}{{|}_{N=1}}:\mathop{{}}_{{}}(\dot{a}-\sqrt{-k})\frac{\partial G(a,f)}{\partial a}=0,~~~~~~~~~~~~~~~~~~~~~
\end{equation}
\begin{equation}
\! \,\,\, {{\delta}_{N}}{{S}_{tot}}{{|}_{N=1}}:\mathop{{}}_{{}}3({{H}^{2}}+\frac{k}{{{a}^{2}}})={{\rho}_{MG}}+{{\rho}_{\sigma}}+{{\rho}_{m}},
\end{equation}

\begin{equation}
{{\delta}_{a}}{{S}_{tot}}{{|}_{N=1}}:\mathop{{}}_{{}}-(2\dot{H}+3{{H}^{2}}+\frac{k}{{{a}^{2}}})={{p}_{MG}}+{{p}_{\sigma}}+{{p}_{m}},
\end{equation}

\begin{equation}
{{\delta}_{\sigma}}{{S}_{tot}}{{|}_{N=1}}:\mathop{{}}_{{}}\frac{1}{2}{f}'(\sigma){{\dot{\sigma}}^{2}}+f(\sigma)(\ddot{\sigma}+3H\dot{\sigma})
+{V}'(\sigma)=0,
\end{equation}
\\
\noindent where $H=\frac{{\dot{a}}}{a}$ is the Hubble parameter,
and ${V}'(\sigma)=\frac{dV(\sigma)}{d\sigma}$.
\\
 In Eqs.~(23)-(26), the effective energy density, and pressure of
dRGT theory are given by

\begin{equation}
{{\rho }_{MG}}=-3m_{g}^{2}\frac{F(a,f)}{{{a}^{3}}},
\end{equation}

\noindent and

\begin{equation}
{{p}_{MG}}=m{{_{g}^{2}}^{{}}}\big[\frac{F(a,f)}{{{a}^{3}}}+(\dot{f}-\beta
)\frac{\partial G(a,f)}{\partial a}\big].
\end{equation}

\noindent respectively. Also, we have

\begin{equation}
{{\rho }_{\sigma }}=\frac{1}{2}f(\sigma){{\dot{\sigma }}^{2}}+V(\sigma ),
\end{equation}
\begin{equation}
{{p}_{\sigma }}=\frac{1}{2}f(\sigma){{\dot{\sigma }}^{2}}-V(\sigma ),
\end{equation}

\noindent for the energy density and pressure of $\sigma$. For action~(2), the equation of motion
 yields

\begin{equation}
{{\delta }_{\varphi }}{{S}_{tot}}{{|}_{N=1}}:\mathop{{}}_{{}}\ddot{\varphi }+3H\dot{\varphi }+\frac{dV(\varphi )}{d\varphi }=0.
\end{equation}

\noindent In this case, the energy density and pressure of
$\varphi $ are also given by
\begin{equation}
{{\rho }_{\varphi }}=\frac{1}{2}{{\dot{\varphi }}^{2}}+V(\varphi ),
\end{equation}

\noindent and
\begin{equation}
_{{}}^{{}}{{p}_{\varphi }}=\frac{1}{2}{{\dot{\varphi
}}^{2}}-V(\varphi ),
\end{equation}

\noindent respectively.

Eq.~(23) has three solutions. For one of these solutions, one obtains
$\dot{a}=\sqrt{-k}$ which leads to $a=\sqrt{-k}t+\text{const}$
that signals us to an open FRW universe as the physical metric
${{g}_{\mu \nu }}$. Similar to flat FRW solution, this solution is
unacceptable. The other two solutions are obtained from
$\frac{\partial G(a,f)}{\partial a}=0$ which leads to

\begin{equation}
{{\beta }_{\pm }}=\frac{1+2{{\alpha }_{3}}+{{\alpha}_{4}}\pm
\sqrt{^{{}}1+{{\alpha }_{3}}+\alpha _{3}^{2}-{{\alpha
}_{4}}}}{{{\alpha }_{3}}+{{\alpha }_{4}}},
\end{equation}

\noindent and

\begin{equation}
a=\frac{\sqrt{-k}f}{{{\beta }_{\pm }}}.
\end{equation}

\noindent Now, assuming $a$ is positive, one finds that, according
to Eq.~(35), ${{\beta}_{\pm }}$ should also be positive. For the
negative scale factor, Eq.~(35) becomes
$a=-\frac{\sqrt{-k}f}{{{\beta}_{\pm }}}$. So, ${{\beta}_{\pm }}$
is always positive. If $f(t)\propto a(t)$, then Eqs.~(27)-(30)
give us

\begin{equation}
\begin{aligned}
& {{\rho }_{MG}}=-{{p}_{MG}}=\Lambda _{\pm }^{{}}= \\
& m_{g}^{2}({{\beta }_{\pm }}-1)[(3-{{\beta }_{\pm }})+{{\alpha }_{3}}(1-{{\beta }_{\pm }})], \\
\end{aligned}
\end{equation}

\noindent where $\Lambda_{\pm }$ denotes the effective
cosmological constant of the theory. Although ${{\alpha}_{3}}$ and
${{\alpha }_{4}}$ are free parameters, Eq.~(34) renders some
constraints. Moreover, if ${{\alpha}_{3}}=-{{\alpha }_{4}}(\pm
(1+{{\alpha }_{3}})>0)$, then the cosmological constant $\Lambda
_{\pm }^{{}}$ becomes infinite. On the other hand, ${{\alpha
}_{4}}={{\frac{3+2{{\alpha }_{3}}+3\alpha _{3}^{2}}{4}}^{{}}}(\pm
(1+{{\alpha }_{3}})>0)$ leads to $\Lambda _{\pm }^{{}}=0$.
Finally, the values of free parameters that lead to negative
radicand in Eq.~(34) are not acceptable (see Ref.~\cite{21}).

\subsection{Inflation}

This subsection examines inflation in the slow-roll
regime for which the scale factor has an exponential behavior, ($a\sim {{e}^{Ht}}(H\sim
\text{const})$), and grows rapidly. Therefore, after a few number
of e-foldings, the term $\frac{k}{{{a}^{2}}}$
becomes ignorable. Hence, Eq.~(24) can be rewritten as

\begin{equation}
3{{H}^{2}}=\frac{1}{2}{f(\sigma)}{{\dot{\sigma }}^{2}}+V(\sigma)+\Lambda _{\pm }^{{}}.
\end{equation}

\noindent Moreover, combining Eq.~(24) with Eq.~(25), one reaches at

\begin{equation}
\dot{H}=-\frac{1}{2}f(\sigma){{\dot{\sigma}}^{2}}.
\end{equation}

\noindent The Hubble slow-roll parameters are also defined as

\begin{equation}
{\varepsilon}_H=-\frac{{\dot{H}}}{{{H}^{2}}},
\end{equation}

\noindent and

\begin{equation}
{{\eta }_{H}}=-\frac{{\ddot{\varphi}}}{H\dot{\varphi}}.
\end{equation}

\noindent In this regard, as the slow-roll conditions, we have
${{\varepsilon }_{H}}, |{\eta}_{H}|<<1$, while the first
condition leads to

\begin{equation}
f(\sigma){{\dot{\sigma}}^{2}}<<V(\sigma)+\Lambda _{\pm }^{{}},
\end{equation}

\noindent meaning that Eq.~(37) can be recast as

\begin{equation}
3{{H}^{2}}\approx V(\sigma)+\Lambda_{\pm}^{{}}.
\end{equation}

\noindent Additionally, for the relation between the first time-derivatives
of $\sigma$ and $\varphi$, we have

\begin{equation}
\dot{\varphi}= \sqrt{f(\sigma)}{\dot{\sigma}}.
\end{equation}

\noindent On the other hand, the second time-derivatives of
$\sigma$ and $\varphi $ address us to

\begin{equation}
\ddot{\varphi}=\sqrt{f(\sigma)}{\ddot{\sigma}}+\frac{1}{2}{{f(\sigma)}^{\frac{-1}{2}}}{f'(\sigma)}{{\dot{\sigma }}^{2}},
\end{equation}

\noindent where $f'(\sigma)=\frac{df}{d\sigma}$. Now, inserting $\dot{\varphi}$ and $\ddot{\varphi }$ in
Eq.~(40), ${{\eta}_{H}}$ is obtained as

\begin{equation}
\eta _{H}=-\frac{\ddot{\sigma}+\frac{1}{2}f(\sigma)^{-1}f'(\sigma) \dot{\sigma }^2}{H \dot{\sigma }}.
\end{equation}

\noindent Using $|{{\eta }_{H}}|<<1,$ Eq.~(26) is approximately
equal to 

\begin{equation}
3H\dot{\sigma }\approx -\frac{1}{f(\sigma)}{V}'(\sigma ).
\end{equation}

Combining Eqs.~(38),~(39), and~(42) with each other, we get

\begin{equation}
{{\varepsilon}_H}=\frac{1}{2}{f(\sigma)}{{(\frac{{\dot{\sigma }}}{H})}^{2}}\approx
\frac{1}{2f(\sigma)}{{\big(\frac{{V}'(\sigma )}{V(\sigma)+\Lambda _{\pm }^{{}}}\big)}^{2}}.
\end{equation}

\noindent and thus

\begin{equation}
3\dot{H}\dot{\sigma }+3H\ddot{\sigma }\approx -\frac{1}{f(\sigma)}
\big[{V}''(\sigma )\dot{\sigma }-{\frac{{f}'(\sigma)}{f(\sigma)}}\dot{\sigma }{V}'(\sigma )\big],
\end{equation}

\noindent in which ${V}''(\sigma )=\frac{{{d}^{2}}V(\sigma
)}{d{{\sigma }^{2}}}$. Finally, it is straightforward to achieve
\begin{equation}
{{\eta }_{H}}\approx \frac{1}{f(\sigma)}\big(\frac{^{{}}{V}''(\sigma )-{\frac{{f}'(\sigma)}{2f(\sigma)}}{V}'{{(\sigma )}^{{}}}}{V(\sigma)+\Lambda _{\pm
}^{{}}}\big)+\frac{{\dot{H}}}{{{H}^{2}}}.
\end{equation}

The best method for investigating a specific potential is the
calculation of  parameters including ${{\varepsilon
}_{V}}$, and ${{\eta }_{V}}$ combined with ${{\varepsilon
}_{V}}\approx {{\varepsilon }_{H}}$, and ${{\eta }_{V}}\approx
{{\eta }_{H}}+{{\varepsilon }_{H}}$ to reach \cite{36}

\begin{equation}
{{\varepsilon }_{V}}=\frac{1}{2f(\sigma)}
{{\big(\frac{{V}'(\sigma )}{V(\sigma)+\Lambda _{\pm }^{{}}}\big)}^{2}},
\end{equation}

\noindent and

\begin{equation}
{{\eta }_{V}}=\frac{1}{f(\sigma)} \big(\frac{^{{}}{V}''(\sigma )-{\frac{{f}'(\sigma)}{2f(\sigma)}}{V}'{{(\sigma )}^{{}}}}{V(\sigma)
+\Lambda _{\pm }^{{}}}\big).
\end{equation}

\noindent ${\varepsilon }_{V}, {|{\eta }_{V}|} << 1$  are the neccessary conditions for the slow-roll inflation \cite{36,37}. The amount of inflation that occurs is measured by the number of e-foldings \cite{38},

\begin{equation}
N_e= \int_{a}^{a_f}d\ln a=\int_{t}^{t_f} {H dt},
\end{equation}

\noindent Thus, using $H dt=\pm \frac{d\sigma}{\sqrt{2{\varepsilon_V}}}$, $N_e$ is rewritten as

\begin{equation}
{N_e}(\sigma)=\left| \int_{\varphi}^{\varphi_f}d\varphi\frac{1}{\sqrt{2{\varepsilon_V}}}\right|=\left|\int_{\sigma}^{\sigma_f}{d\sigma} f(\sigma)\big(\frac{V(\sigma )+\Lambda_{\pm }}{{V}'(\sigma )}\big)\right|.
\end{equation}
\\
\noindent  In Eqs.~(52) and~(53), the subscript $f$ point to the end of inflation \cite{38}. The scalar spectral index $n_s$, and tensor-to-scalar ratio $r$ are thus calculated as

\begin{equation}
 n_s=1-6{{\varepsilon }_{V}}+2{{\eta }_{V}},
\end{equation}
\noindent and
\begin{equation}
 r=16{{\varepsilon }_{V}},
\end{equation}

\noindent respectively. Here, $r$ is a significant observational
quantity, and it is used to distinguish different inflationary
models.

\section{NUMERICAL DYNAMICS}

This section focuses on the specific function $f(\sigma)={A}{{|\sigma|}^{-p}} (p\not =0)$ which has pole at $\sigma =0$ for $p>0$ \cite{25,26,27,28,29}. Because the field does not cross zero due to the pole, $\sigma$ can be lie within one of the branches $\sigma > 0$ or $\sigma < 0$. The case $p=2$, and $V(\sigma)=0$ corresponds to minimal $k$-essence model \cite{29}.\\
\noindent Using Eq.~(4), the relationships between $\sigma$ and $\varphi$ are easily achieved as 

\begin{equation}
\varphi =\left\{ \begin{aligned}
& \frac{2\sqrt{A}}{2-p}{{|\sigma|}^{\frac{2-p}{2}}}~~~\text{for}~~p\ne 2\underset{\underset{{}}{\mathop {}}\,}{\mathop {}}\, \\
&~~~~~~~~~~~~~~~~~~~~~~~~~~~~~~~~~~~~,\\
& \sqrt{A}\ln(|\sigma|) ~~~~~\text{for}~~p=2 \\
\end{aligned} \right.
\end{equation}

\noindent and 

\begin{equation}
\sigma =\left\{ \begin{aligned}
& \pm \big({\frac{(2-p)\varphi}{2\sqrt{A}}}\big)^{\frac{2}{2-p}}~~~\text{for}~~p\ne 2\underset{\underset{{}}{\mathop {}}\,}{\mathop {}}\, \\
&~~~~~~~~~~~~~~~~~~~~~~~~~~~~~~~~~~~~~~~~~~~,\\
& \pm e^{\frac{\varphi}{\sqrt{A}}} ~~~~\,\qquad \qquad\text{for}~~p=2 \\
\end{aligned} \right.
\end{equation}

\noindent respectively, up to an integration constant.

\subsection{The  Power Potential}

First, let us consider $V(\sigma)={V_0}{{\sigma }^{n}}$, where $V_0$ and $n$ are non-zero constants. The general form of canonicalized potential is obtained using Eq.~(57) as

\begin{equation}
V(\varphi) =\left\{ \begin{aligned}
& \pm {V_0} \big({\frac{(2-p)\varphi}{2\sqrt{A}}}\big)^{\frac{2n}{2-p}}~~~\text{for}~~p\ne 2\underset{\underset{{}}{\mathop {}}\,}{\mathop {}}\, \\
&~~~~~~~~~~~~~~~~~~~~~~~~~~~~~~~~~~~~~~~~~~~~~~.\\
& \pm{V_0} e^{\frac{n\varphi}{\sqrt{A}}} ~~~\,\,\qquad\quad\, \quad\text{for}~~p=2 \\
\end{aligned} \right.
\end{equation}

\noindent Inserting ${V_0}{{\sigma }^{n}}$ in
Eq.~(50), and using Eq.~(51), $\varepsilon_V$ and
$\eta_V$ are calculated as

\begin{equation}
\varepsilon_V=\frac{{|\sigma|}^{p}}{2A}{{\big(\frac{n{{\sigma }^{n-1}}}{{{\sigma}^{n}}+c}\big)}^{2}},
\end{equation}

\noindent and

\begin{equation}
\eta _V=\frac{{|\sigma|}^{p}}{2A}\big[\frac{^{{}}n(2n+p-2){{\sigma
}^{n-2}}^{{}}}{{{\sigma }^{n}}+c}\big],
\end{equation}

\noindent respectively, where $c$ is defined as $c=\frac{{\Lambda}_{\pm }}{V_0}$. As inflation ends when ${\varepsilon_V}=|{\eta_V}|=1$ \cite{39,40}, we have

\begin{equation}
{\tilde{\sigma}}_f={\big(\frac{2n+p-2}{-3n-p+2}\big)}^{\frac{1}{n}},
\end{equation}
\noindent and

\begin{equation}
A=\frac{1}{2}{{\big(c\frac{2n+p-2}{-3n-p+2}\big)}^{\frac{p-2}{n}}}{{(2n+p-2)}^{2}},
\end{equation}

\begin{figure*}
\begin{center}
\subfloat{\includegraphics[width=0.5\linewidth]{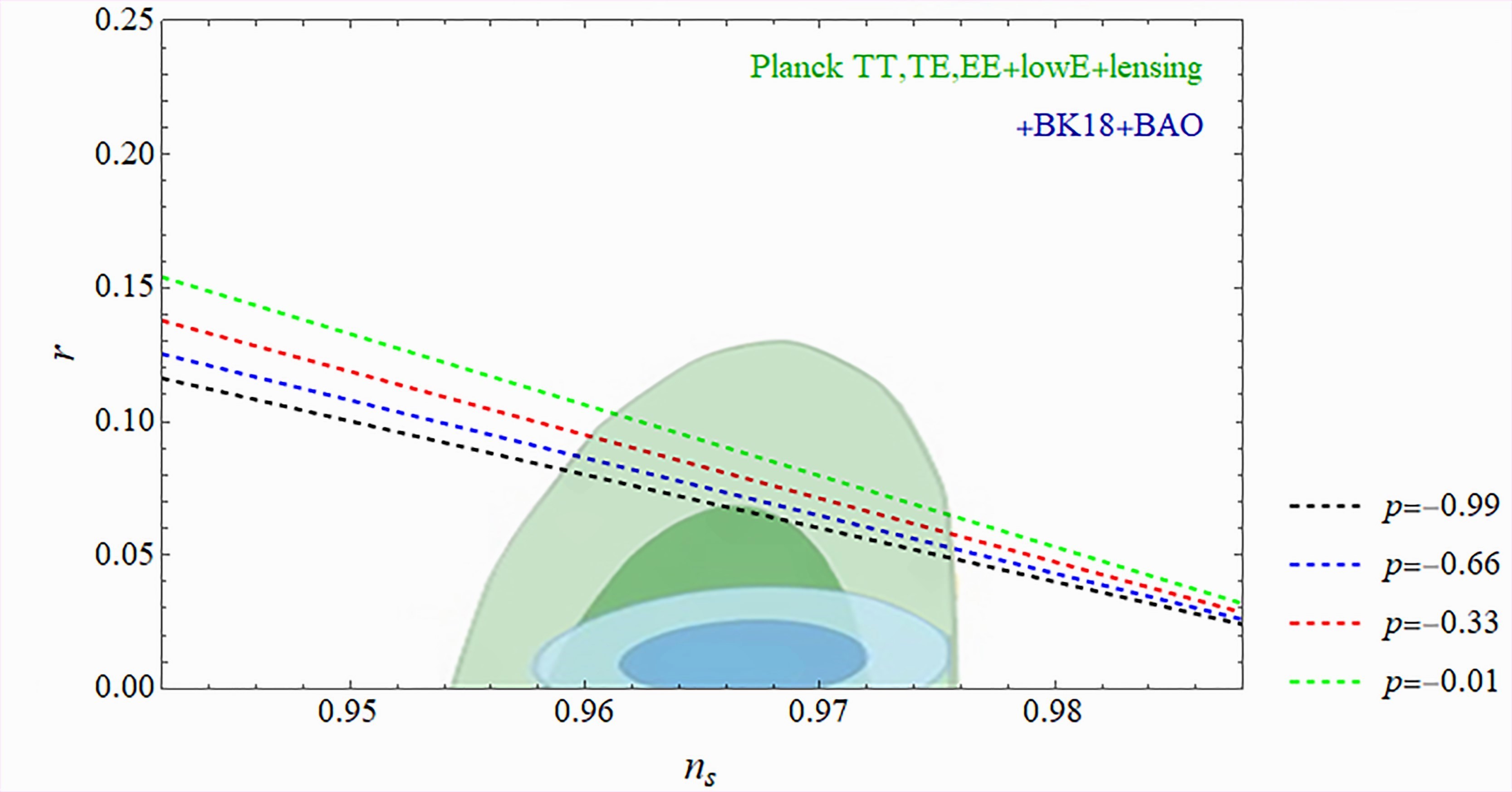}}
\subfloat{\includegraphics[width=0.5\linewidth]{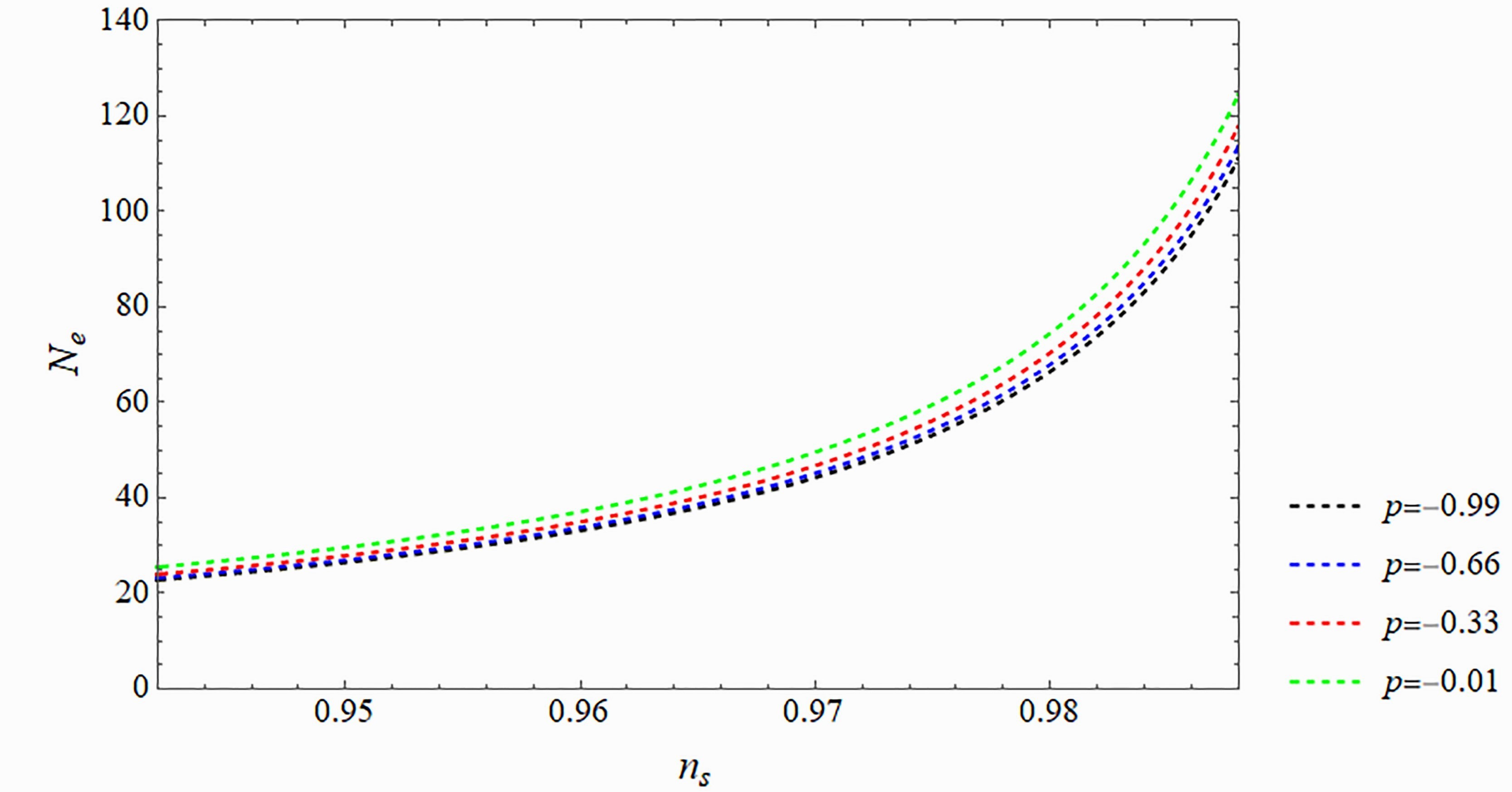}}\\
\subfloat{\includegraphics[width=0.5\linewidth]{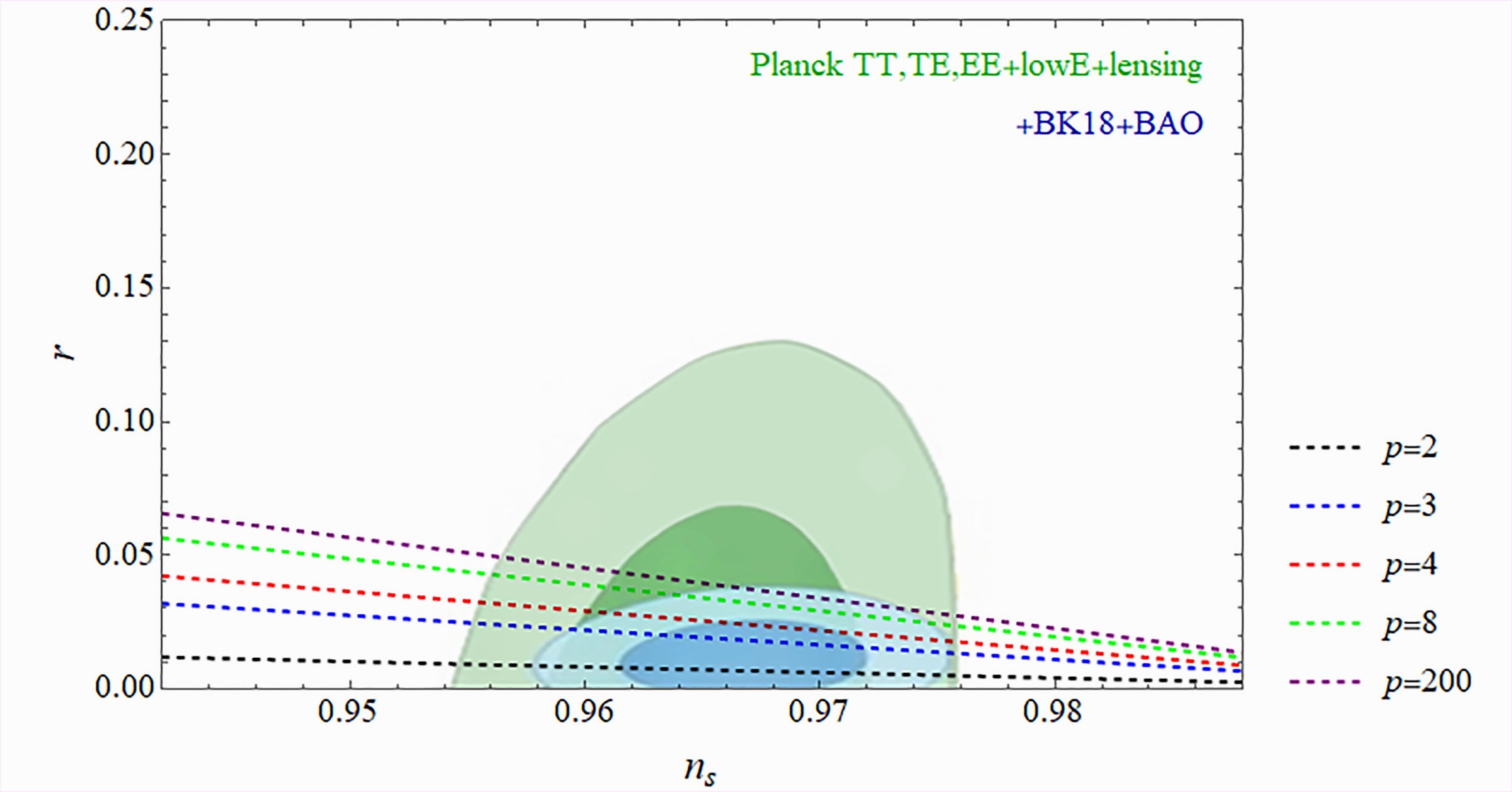}}
\subfloat{\includegraphics[width=0.5\linewidth]{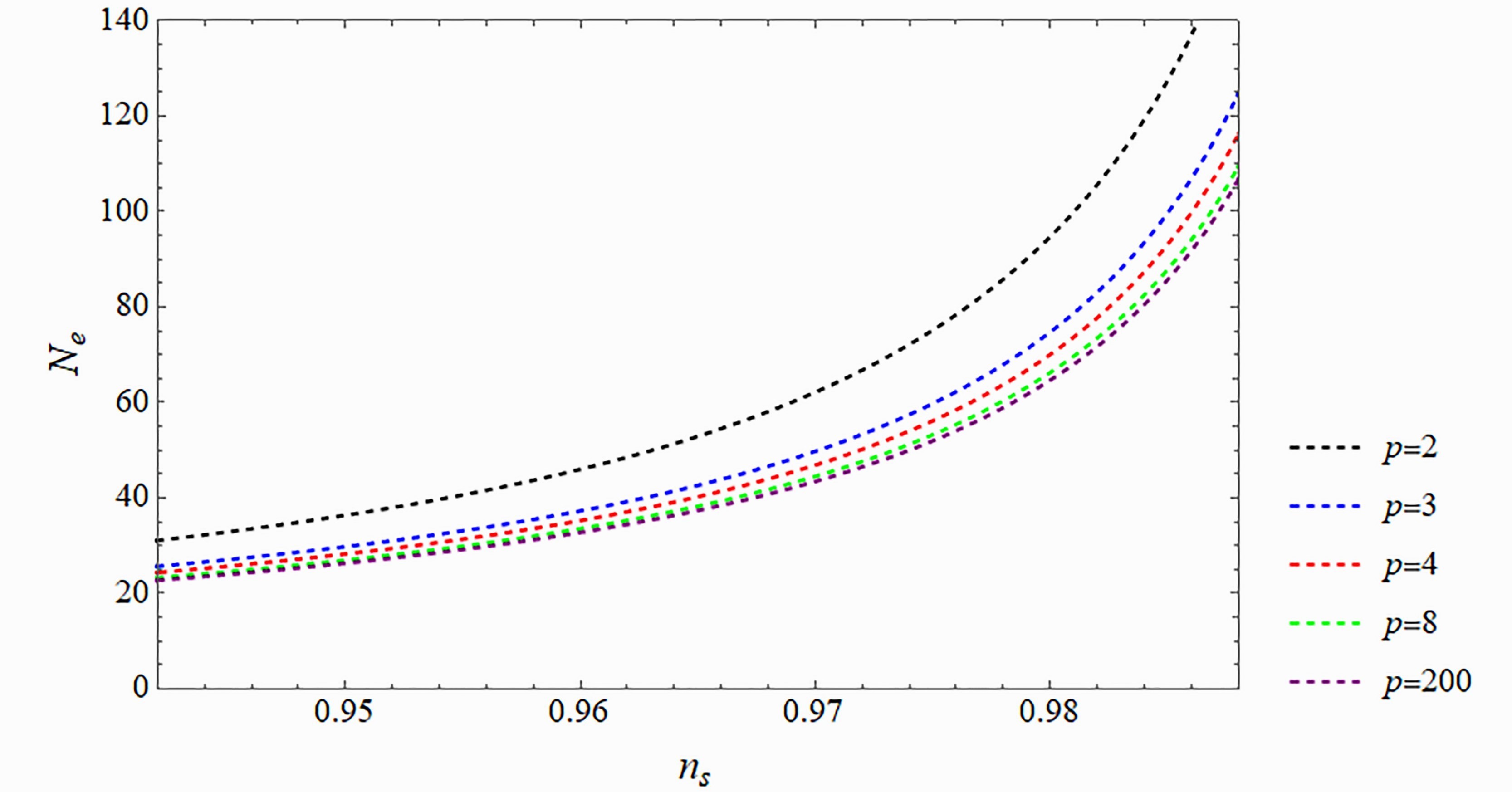}}\\
\caption{\footnotesize Plots  of $r(n_s)$ and ${N_e}(n_s)$ for $V_0 \sigma$ ($n=1$) with ${V_0}>0$ (top plots) and ${V_0}<0$ (bottom plots). In left plots, green and blue regions are related to
			the Planck 2018 data \cite{34} and its combination with BK18 and BAO \cite{35}, respectively. In addition, the 68\% conf-\\idence
			regions are distinguished from the 95\% confidence regions by
			highlighting the corresponding regions. Depending on the values of\\ $p$, the curves of $r(n_s)$ for ${V_0}>0$ lie within the 68\% or 95\% confidence regions of the Planck 2018 TT,TE,EE+lowE+lensing data, wher-\\eas the curves of $r(n_s)$ for ${V_0}<0$ and $2<p<200$ lie within the 68\% confidence region of the Planck 2018 TT,TE,EE+lowE+lensing+\\BK18+BAO data. More details on the text.}\label{fig1}	
\end{center}					
\end{figure*} 

\noindent where $\tilde{\sigma}\equiv{\sigma}{c^{\frac{-1}{n}}}$. Only $n=1$ is acceptable for $\sigma < 0$ because other values of $n$ cause $\sigma_f$ or $V(\sigma)$ to become imaginary numbers, thus the branch $\sigma > 0$ is considered here. In table.~(\rom{1}), three ranges of $p$ for each $n$ are obtained using condition $\sigma_f > 0$, whereas ${V_0} > 0$ in case 1 and ${V_0} < 0$ in cases 2 and 3. It should be noted that $p=-3n+2$ and $p=-2n+2$ are unacceptable because they lead to $A=0$ or $A=\infty$, causing undifinable expressions for $\varphi$~(56) and the canonicalized potential~(58). Moreover, in this manner, the kinetic term in the Lagrangian~(1) becomes zero or infinite. Consequently, we do not pay attention to these improper solutions. Case 3 leads to $ V(\sigma)+\Lambda_{\pm}^{{}} < 0$, indicating that the slow-roll condition~(41) has been violated \cite{41}. Hence, this case is not also studied in the following. 

\begin{table}[htp!]
\renewcommand*{\arraystretch}{1.6}
\scalebox{0.86}{ %
\begin{tabular}{|C{40pt}|C{120pt}|C{120pt}|}
\hline
Case & Range of $p$ for $n > 0$ & Range of $p$ for $n < 0$ \\
\hline
1 & $-3n+2 < p < -2n+2$ & $-2n+ 2 < p < -3n+2$ \\
\hline
2 & $-2n+2 < p$ &$-2n+2 > p$ \\
\hline
3 & $-3n+2 > p$ &$-3n+2 < p$ \\
\hline	
\end{tabular}}
\caption{\footnotesize Allowed ranges of $p$ depending value of $n$ whereas ${V_0} > 0$ in case 1 and ${V_0} < 0$ in cases 2 and 3.} \label{tab1}
\end{table}

\noindent Focusing on cases 1 (${V_0}>0$) and 2 (${V_0}<0$), two ranges of $n$ (named $i$ and $j$) for which the Lagrangian (1) may include pole gathered in Table.(\rom{2}). Indeed, while we have always $p>0$ in the range $i$, both of the positive and negative values of $p$ are allowed in range $j$.

\vspace{0.75cm}
\begin{table}[htp!]
\renewcommand*{\arraystretch}{1.6}
\begin{tabular}{|C{40pt}|C{100pt}|C{100pt}|}
\hline
Case & Range $i$ & Range $j$ \\
\hline
1 &  $0 <n \leq \frac{2}{3}$ & $\frac{2}{3} < n < 1$\\
\hline
2 & $1 \geq n$ & $1 < n$\\
\hline
\end{tabular} 
\caption{\footnotesize The ranges of $n$ which lead to $p > 0$ and thus pole.}\label{tab2}
\end{table}

\noindent Inserting $V(\sigma )=V_0{{\sigma
}^{n}}$ in Eq.~(53), the number of e-foldings $N_e$ is obtained as

\begin{equation}
{N_e}(\sigma)=\left|\frac{{A}'}{n}\int_{{\tilde{\sigma}}}^{{\tilde{\sigma}}_f} {d\tilde{\sigma}}
({{\tilde{\sigma}}^{-p+1}}+{{\tilde{\sigma}}^{-p-n+1}})\right|,
\end{equation}

\noindent in which

\begin{equation}
\int_{{\tilde{\sigma}}}^{{\tilde{\sigma}}_f} {d\tilde{\sigma}}{{\tilde{\sigma}}^{-p+1}}=\left\{ \begin{aligned}
 & \frac{1}{-p+2}({{\tilde{\sigma}}_f}^{-p+2}-{{\tilde{\sigma}}}^{-p+2})~~\,\text{for}~~p\ne 2 \\
&~~~~~~~~~~~~~~~~~~~~~~~~~~~~~~~~~~~~~~~~~~~~~~~~ ,
 \\                                                                    
 & \ln(\frac{{\tilde{\sigma}}_f}{{\tilde{\sigma}}}) ~~~~~~~~~~~~~~~~~~~~~~~~~\text{for}~~p=2 \\
\end{aligned} \right.
\end{equation}
\\
\begin{figure*}[htp]
\subfloat{\includegraphics[width=0.5\linewidth]{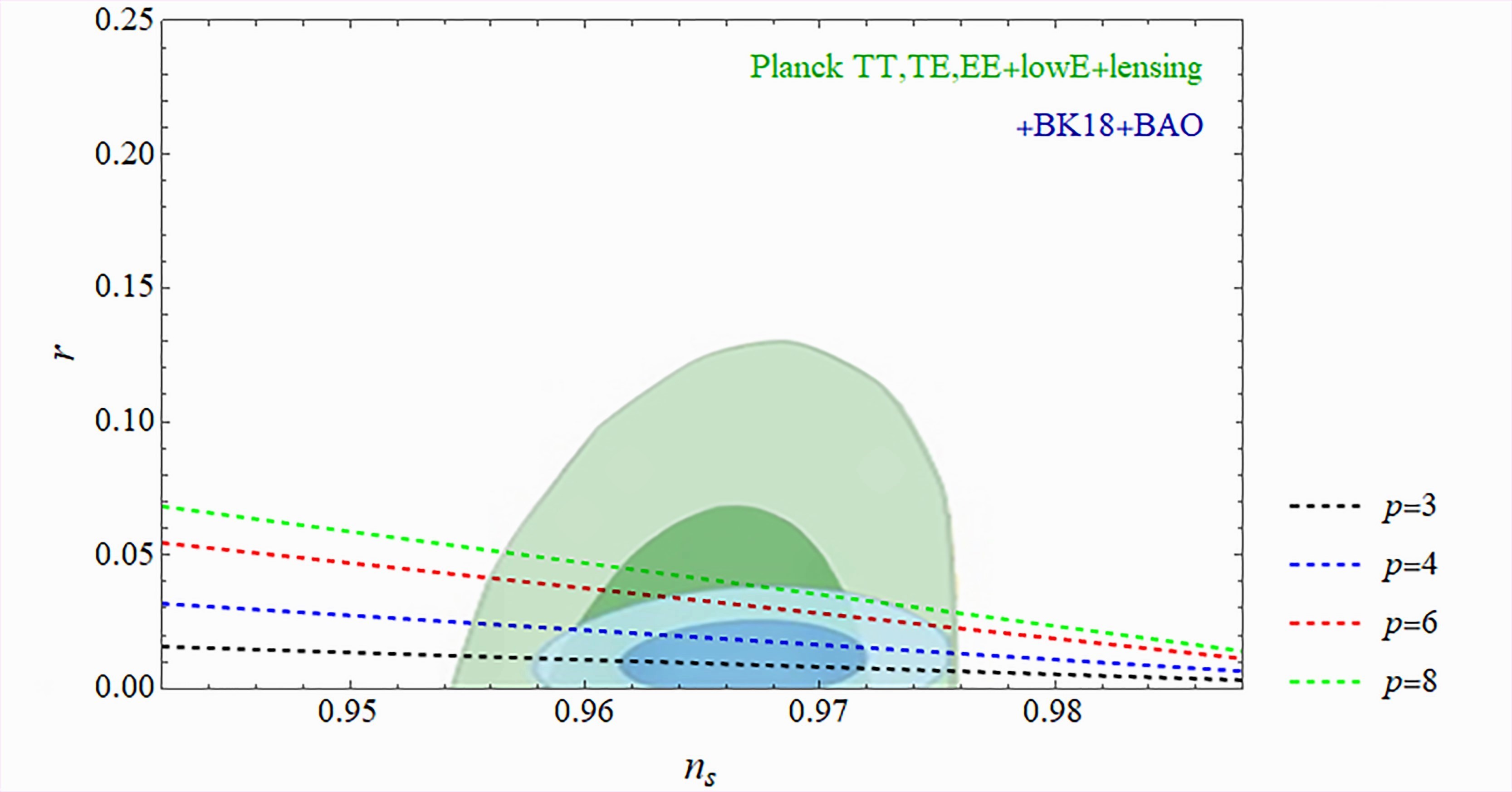}}
\subfloat{\includegraphics[width=0.5\linewidth]{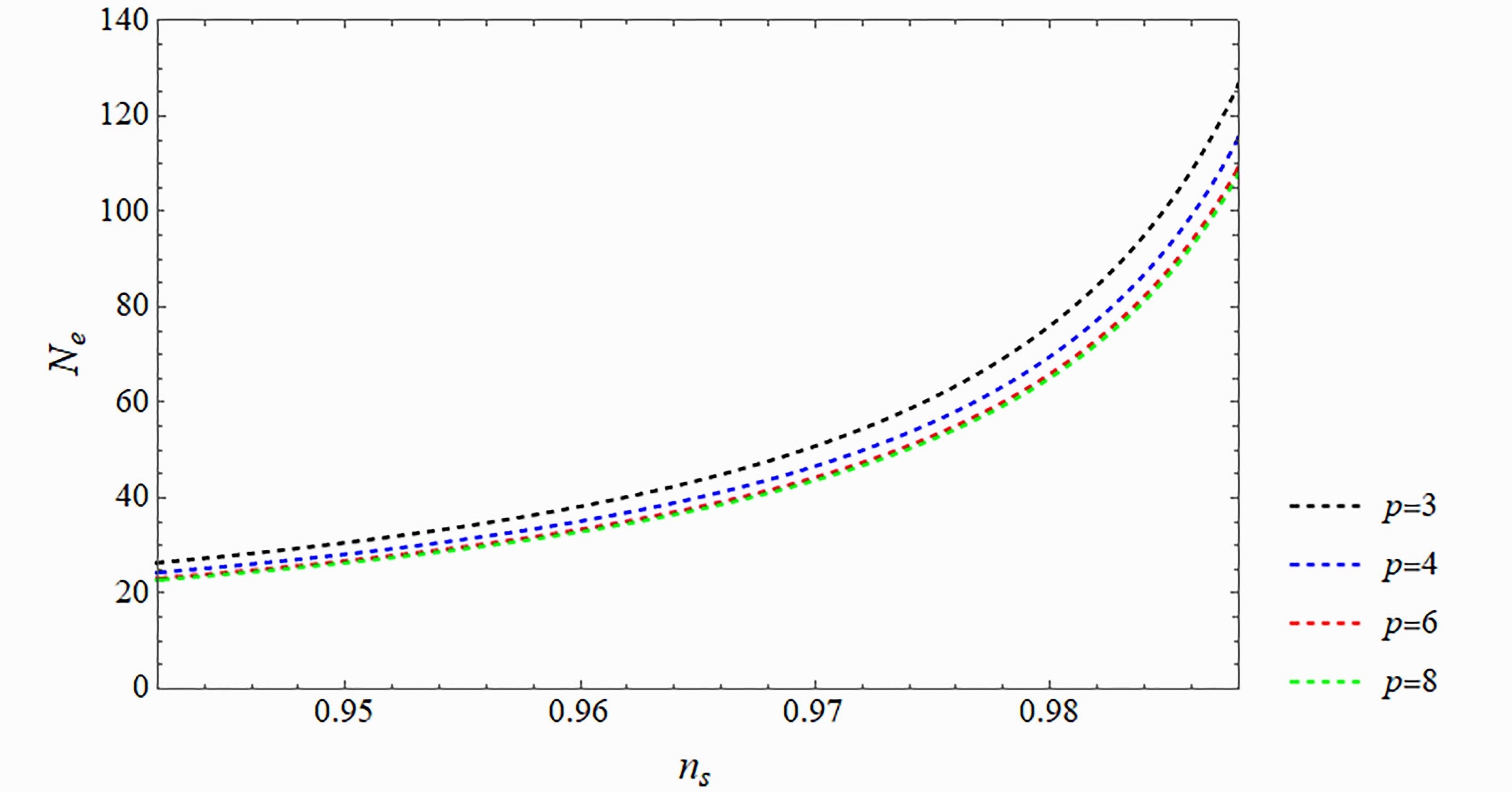}}\\
\subfloat{\includegraphics[width=0.5\linewidth]{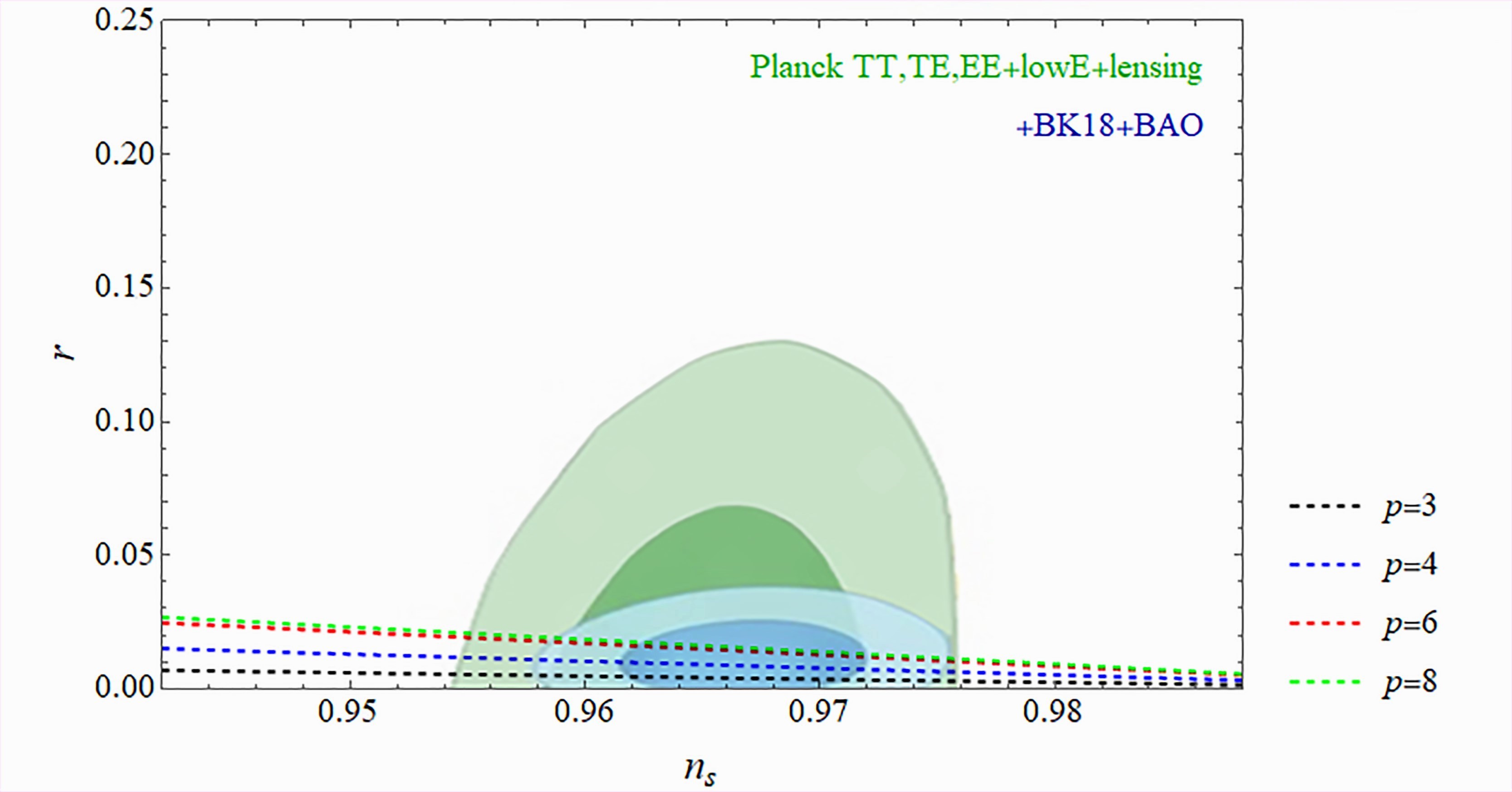}}
\subfloat{\includegraphics[width=0.5\linewidth]{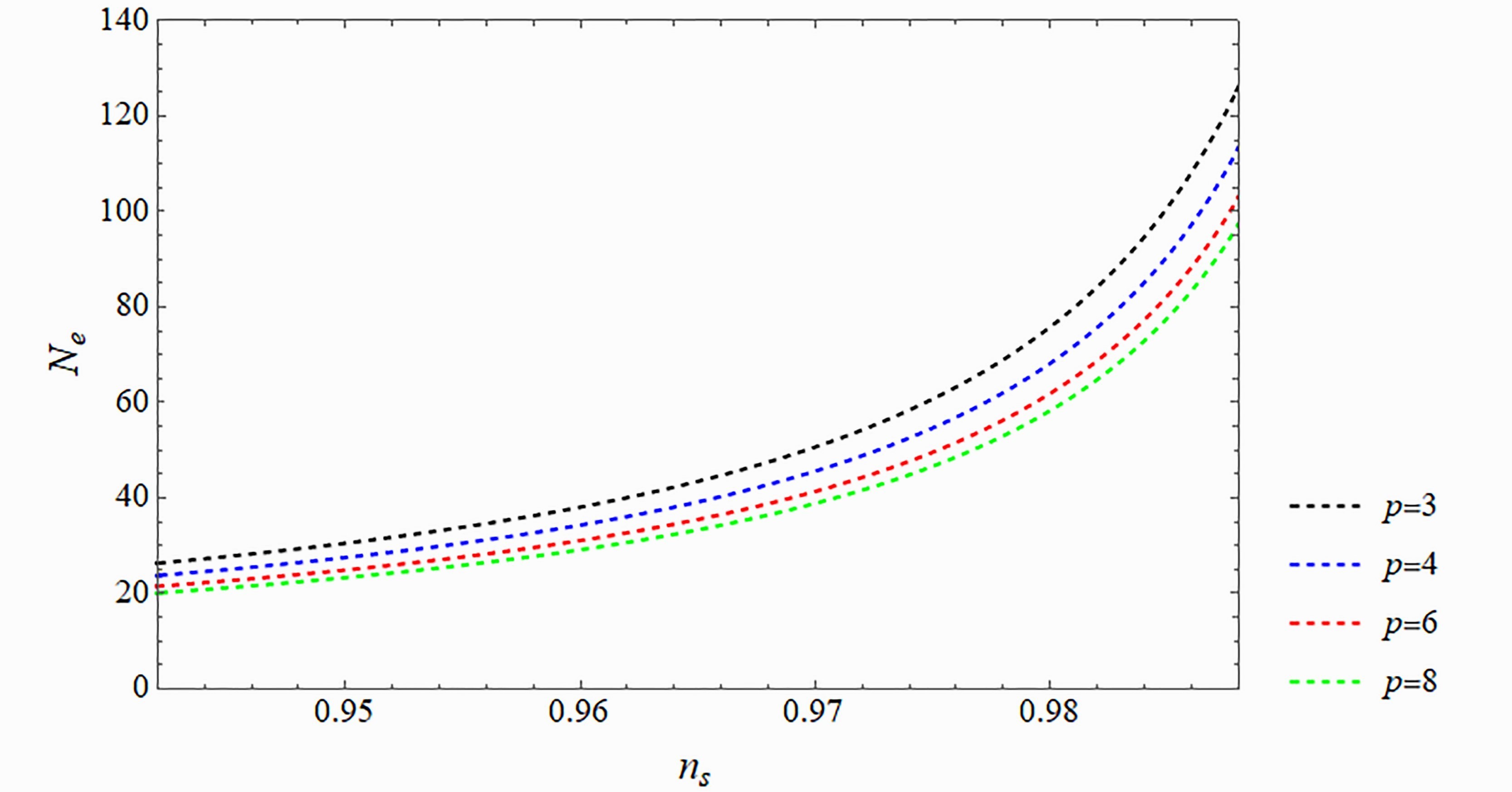}}\\
\caption{\footnotesize Plots of $r(n_s)$ and ${N_e}(n_s)$ for the
        exponential potential $V_0 e^{-\tilde{\sigma}}$, $p>0$ and $c=0$ (top plots) and $c=0.5$ (bottom plots). In left plots, green and blue regions are related to
        the Planck 2018 data \cite{34} and its combination with BK18 and BAO \cite{35}, respectively. In ad-\\dition, the 68\% confidence regions are highlighted compared to the 95\% confidence regions. Depending on the values of $p$, the curves of $r(n_s)$ for $c=0$ lie within the 68\% or 95\% confidence regions of the Planck 2018 TT,TE,EE+lowE+lensing+BK18+BAO data. In addit-\\ion, all curves of $r(n_s)$ for $c=0.5$ lie within the 68\% confidence region of the Planck 2018 TT,TE,EE+lowE+lensing+BK18+BAO data. More details on the text.}\label{fig2}       
\end{figure*}

\noindent and

\begin{equation*}
\hspace{-5cm}\int_{\tilde{\sigma}}^{\tilde{\sigma}_f}{d\tilde{\sigma}}{{\tilde{\sigma}}^{-p-n+1}}=
\end{equation*}
\begin{equation}
\left\{\begin{aligned}
& \frac{1}{-p-n+2}({{\tilde{\sigma}}_f}^{-p-n+2}-{{\tilde{\sigma}}}^{-p-n+2})~~~~\text{for}~~p\ne 2\underset{\underset{{}}{\mathop {}}\,}{\mathop {}}\, \\
&~~~~~~~~~~~~~~~~~~~~~~~~~~~~~~~~~~~~~~~~~~~~~~~~~~~~~~~~~~~~~~,\\ 
& \ln(\frac{{\tilde{\sigma}}_f}{\tilde{\sigma}}) ~~~~~~~~~~~~~~~~~~~~~~~~~~\qquad \qquad \,\,\text{for}~~p=2 \\
\end{aligned} \right.
\end{equation} 

\noindent where $A'\equiv c^{\frac{2-p}{n}}{A}$. Now, equipped with Eqs. (54), (55), (59), and (60), the
scalar spectral index $n_s$, and tensor-to-scalar ratio $r$ are
obtained as

\begin{equation}
n_s=1+n\frac{{\tilde{\sigma}}^{p-2}}{{A}'}\big[\frac{(-n+p-2)+(2n+p-2){{\tilde{\sigma}}^{-n}}}{{{(1+{{\tilde{\sigma}}^{-n}})}^{2}}}\big],
\end{equation}
\noindent and
\begin{equation}
r=8\frac{{\tilde{\sigma}}^{p-2}}{{A}'}\big({\frac{n}{1+{{\tilde{\sigma}}^{-n}}}}\big)^{2},
\end{equation}

\noindent respectively. From the given equations, two striking conclusions can be drawn. First, $r(n_s)$ and ${N_e}(n_s)$ are independent of $c\,(c\not=0)$ because $\sigma_f$ and $\sigma$ are proportional to $c^{\frac{1}{n}}$ for each specific $n$ and $p$. Secondly, the {curves} of $r(n_s)$ and ${N_e}(n_s)$ are repeatable for any values of $n$, depending on the values of $p$. In this regard, three instances for each of cases 1 and 2 are given in table.~(\rom{2}). Note that for $c=0$,~$p=-3n+2$,~$A$ becomes a free parameter, and therefore the second conclusion is rendered invalid.

Using Eq. (66), $\sigma$ is obtained for $0.942<n_s<0.988$. The values of $r(n_s)$ and ${N_e}(n_s)$ are then calculated by substituting them into Eqs. (63) and (67). In Fig. (\ref{fig1}), $r(n_s)$ and ${N_e}(n_s)$ are shown for 
$V(\sigma)=V_0 \sigma$, with top and bottom plots corresponding to cases 1 and 2, respectively. Depending on the values of $p$, the curves of $r(n_s)$ for case 1 lie within the 68\% or 95\% confidence regions of the Planck 2018 TT,TE,EE+lowE+lensing data, whereas the curves of $r(n_s)$ for case 2 and $2<p<200$ lie within the 68\% or 95\% confidence regions of the Planck 2018 TT,TE,EE+lowE+lensing+BK18+
BAO data. For $p>200$, the curves of $r(n_s)$ increase very gradually, get away from the Planck 2018 TT,TE,EE+lowE+lensing+BK18+BAO data, and fall into the 68\% confidence region of the Planck 2018 TT,TE,EE+lowE+lensing data. Furthermore,~${N_e}(n_s)$ is highly similar to $p=200$ for $p>200$. As previously stated, only $\sigma>0$ is admissible and correspondingly $p<2$, leading to $\sigma<0$ is ignored.

\begin{figure*}[htp]
\subfloat{\includegraphics[width=0.5\linewidth]{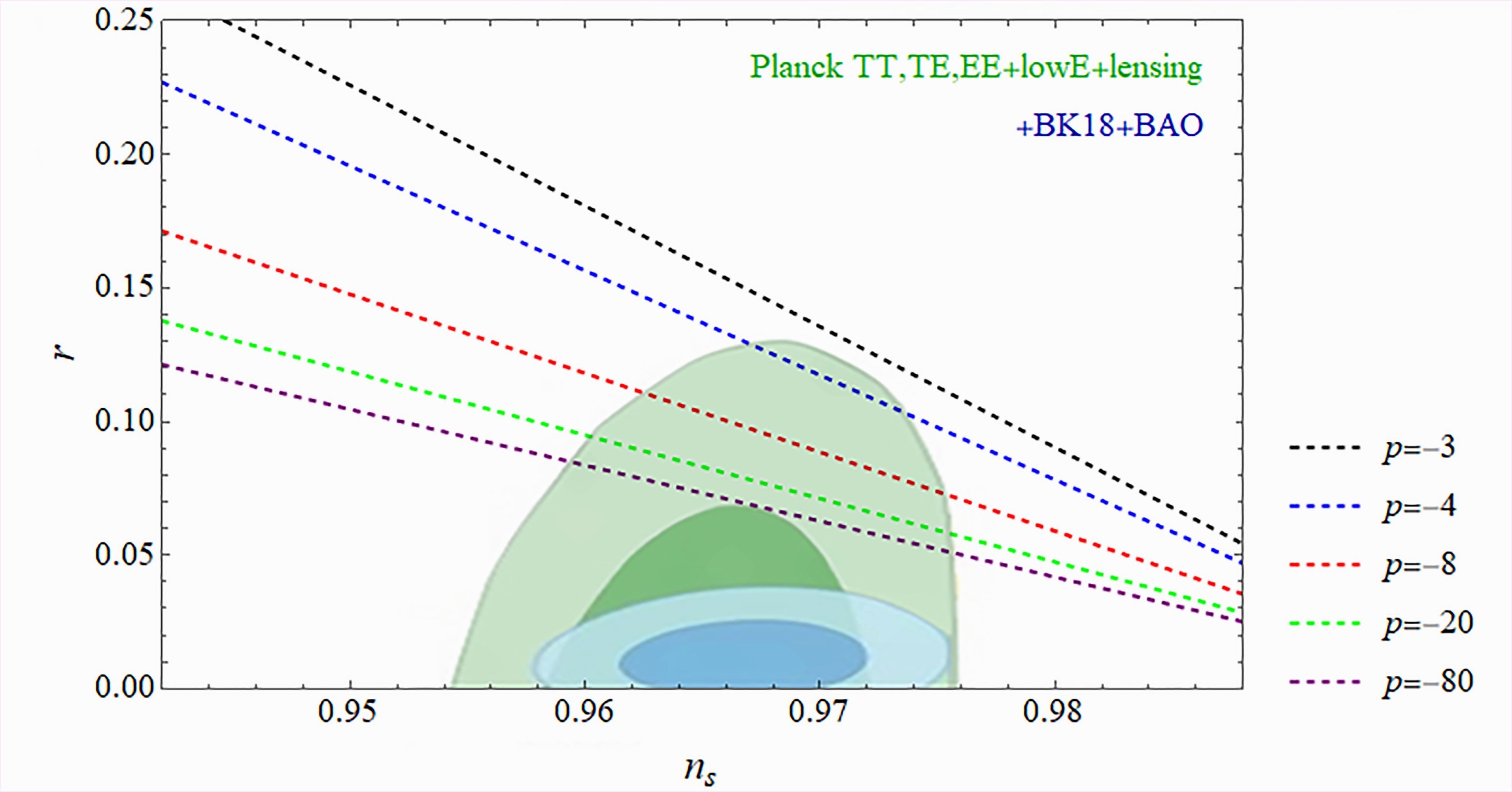}}
\subfloat{\includegraphics[width=0.5\linewidth]{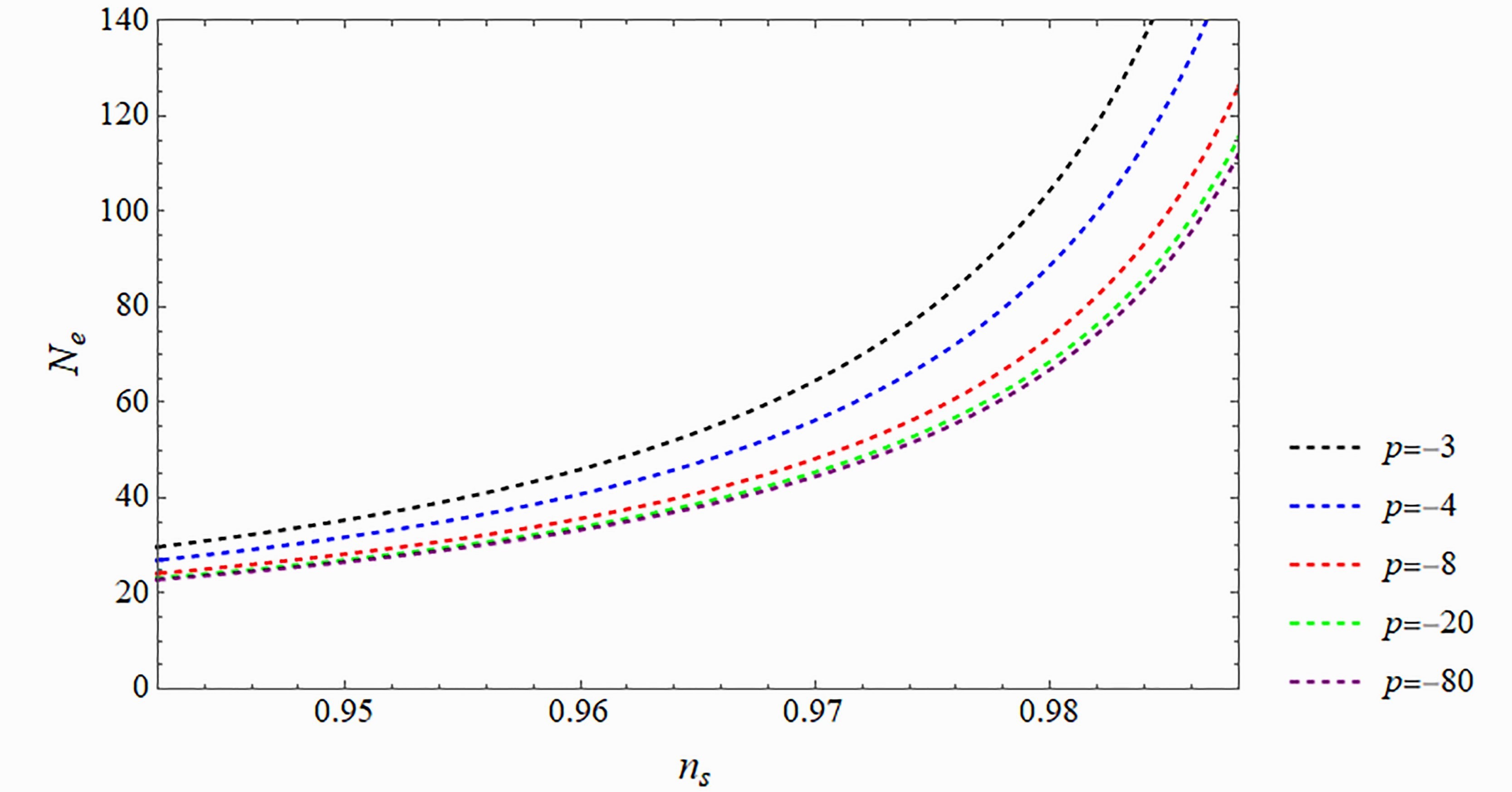}}\\
\subfloat{\includegraphics[width=0.5\linewidth]{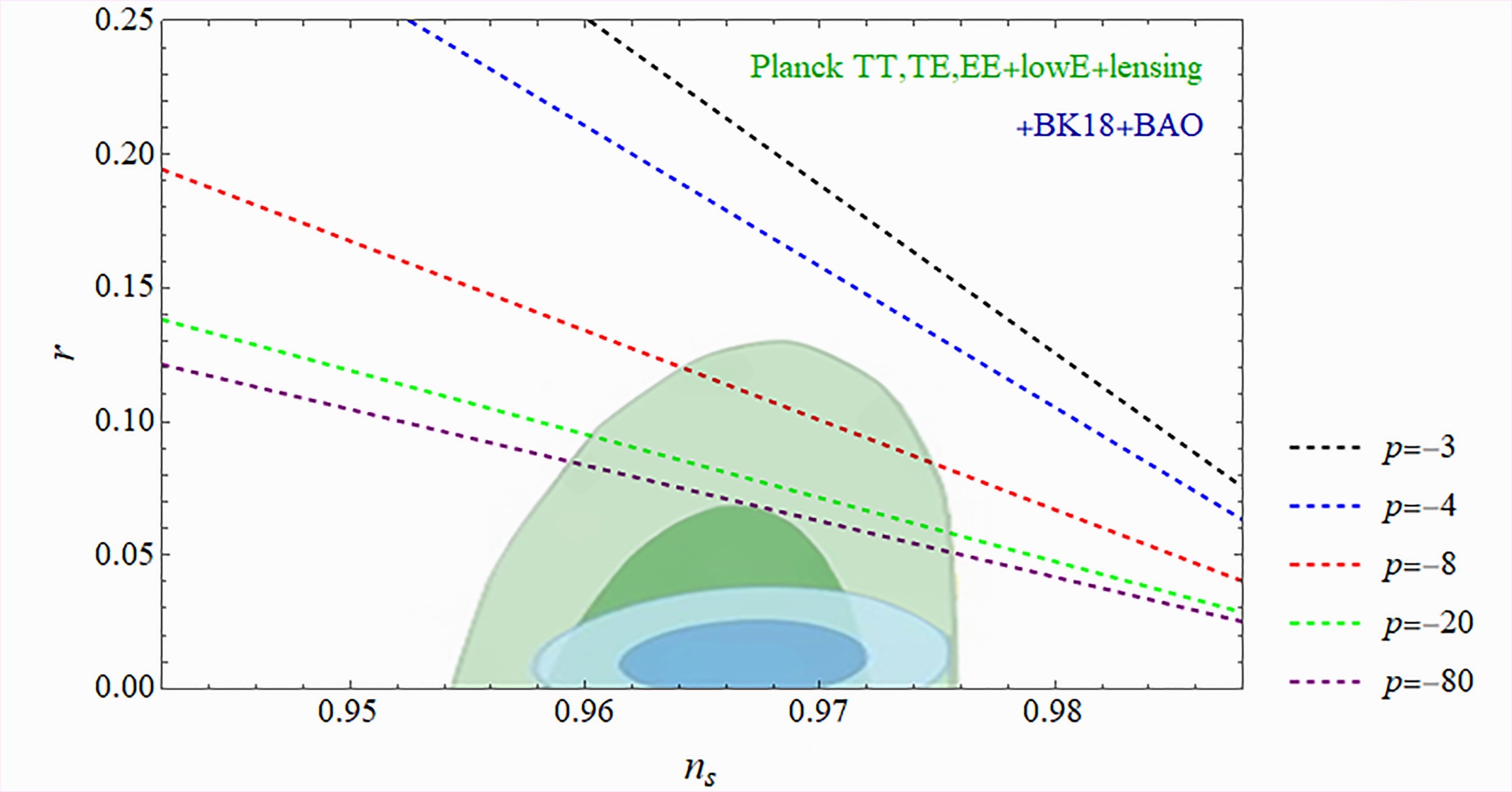}}
\subfloat{\includegraphics[width=0.5\linewidth]{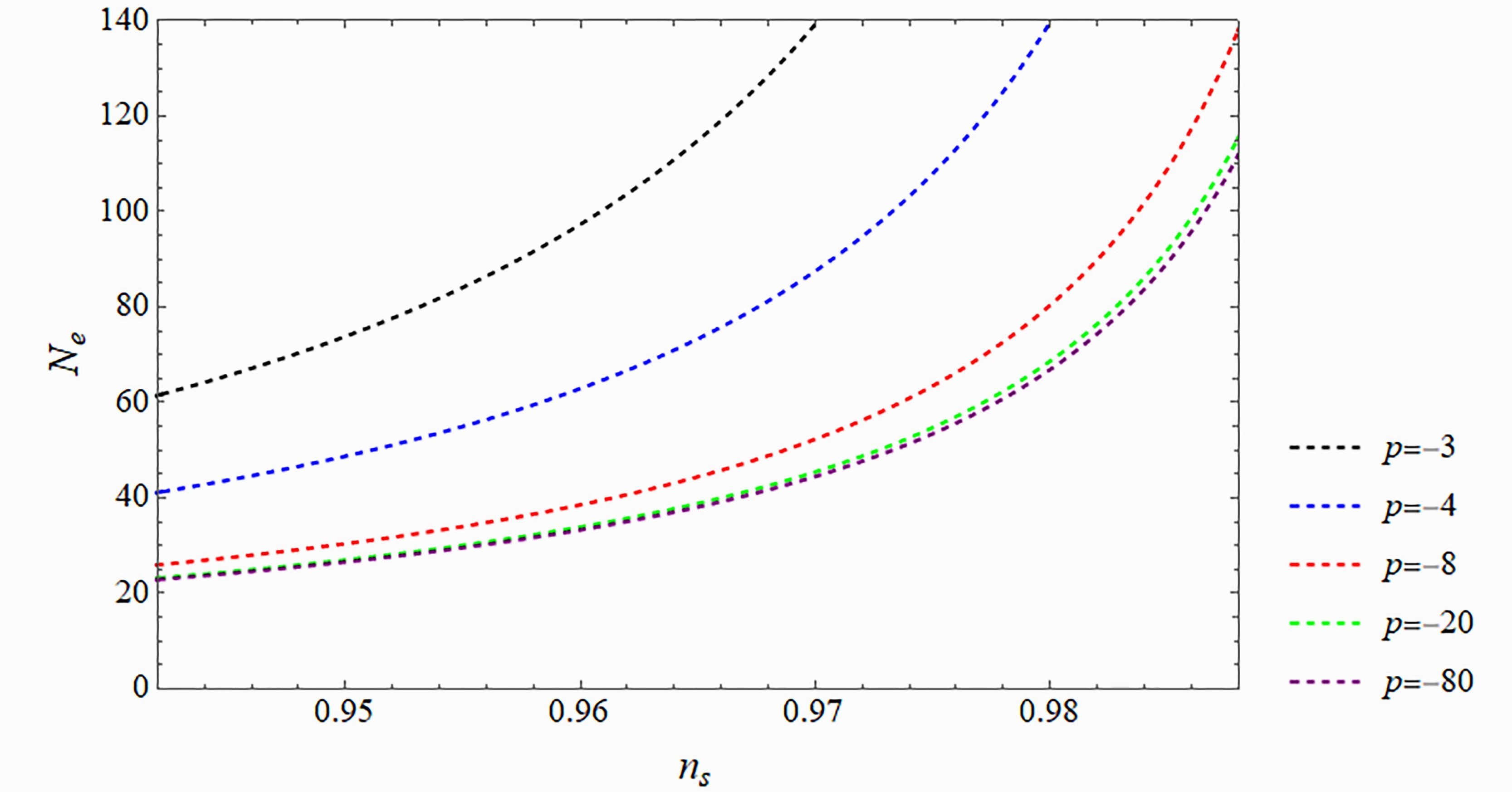}}
\caption{\footnotesize Plots of $r(n_s)$ and ${N_e}(n_s)$ for the
        	exponential potential $V_0 e^{-\tilde{\sigma}}$. Here, $p<0$ and $c=0$ (top plots) and $c=10$ (bottom plots). In left plots, green and blue regions are related to
        	the Planck 2018 data \cite{34} and its combination with BK18 and BAO \cite{35}, respectively. In addition,
        	the 68\%confidence regions are highlighted compared to the 95\% confidence regions. Depending on the values of $p$ and $c$, the cur-\\ves of $r(n_s)$ lie within the 68\% or
        	 95\%confidence regions of the Planck 2018 TT,TE,EE+lowE+lensing data, in such a way that $c=0$ is obviously more consistent with observational data than $c=10$. More details on the text.}\label{fig3}        	        	 
\end{figure*}

\begin{table}[htp!]
\renewcommand*{\arraystretch}{1.6}
\begin{tabular}{|C{55pt}|C{65pt}|C{120pt}|}
\hline
Case& Power $n$ & Range of $p$ \\
\hline
  \multirow{3}{*}{1}  & 1 & $-1 < p < 0$\\
    \cline{2-3}
& -1 & $4 < p < 5$\\
    \cline{2-3}
    & 2 & $-4 < p < -2$\\
\hline
 \multirow{3}{*}{2}  & 1 & $0 < p$\\
    \cline{2-3}
  & -1 & $4 > p$\\
    \cline{2-3}
   & 2 & $2 < p$\\
\hline
\end{tabular}
\caption{\footnotesize Different values of $n$ and their corresponding intervals of $p$ generating the same results as those of Fig.~(\ref{fig1}).} \label{tab3}
\end{table}
\subsection{The Exponential Potential}

In this subsection, we study the inflationary era generated by the
exponential potential $V(\sigma)=V_0 e^{-\gamma \sigma}$ where
$\gamma$ is a constant. Using Eq.~(57), the general form of canonicalized potential is achieved as

\begin{equation}
V(\varphi) =\left\{ \begin{aligned}
&{V_0} e^{\pm \gamma {\big(\frac{(2-p)\varphi}{2\sqrt{A}}\big)}^{\frac{2}{2-p}}} ~~~\text{for}~~p\ne 2\underset{\underset{{}}{\mathop {}}\,}{\mathop {}}\, \\
&~~~~~~~~~~~~~~~~~~~~~~~~~~~~~~~~~~~~~~~~~~~.\\
& {V_0} e^{\pm \gamma e^{\frac{\varphi}{\sqrt{A}}}} ~~~~~~~~~~~\,\,\,\,\text{for}~~p=2 \\
\end{aligned} \right.
\end{equation}

\noindent Inserting
$V_0 e^{-\gamma \sigma}$ in Eq.~(50), and Eq.~(51), the slow-roll
parameters are obtained as

\begin{equation}
{\varepsilon }_{V}=\frac{|\sigma|^p}{2A}\big(\frac{\gamma}{1+ce^{\gamma \sigma}}\big)^2,
\end{equation}

\noindent and

\begin{equation}
{\eta }_{V}=\frac{|\sigma|^p}{A}\big(\frac{\gamma(\gamma-\frac{p}{2}\sigma^{-1})}{1+ce^{\gamma \sigma}}\big).
\end{equation}

\noindent At the point of ${\varepsilon }_{V}=|{\eta }_{V}|=1$,
inflation is ended, a fact helps us find out

\begin{equation}
c(e^{{\tilde{\sigma}}_f})=\frac{-3{{\tilde{\sigma}}_f}+p}{2{{\tilde{\sigma}}_f}-p},
\end{equation}

\noindent and

\begin{equation}
A=\gamma^{2-p}\big[\frac{1}{2}{{|{\tilde{\sigma}}_f|}^{p-2}}(2{{\tilde{\sigma}}_f}-p)^{2}\big],
\end{equation}

\noindent where $\tilde{\sigma}$ is defined as $\tilde{\sigma}=\gamma \sigma$. Inserting $V_0 e^{-\tilde{\sigma}}$ in Eq.~(53), the number of e-foldings $N_e$ are obtained as

\begin{equation}
{N_e}(\sigma)=\left|A'\int_{\tilde{\sigma}}^{{\tilde{\sigma}}_f} d\tilde{\sigma}\big[{{|{\tilde{\sigma}}|}^{-p}}(1+ce^{\tilde{\sigma}})\big]\right|,
\end{equation}

\noindent in which

\begin{equation*}
\hspace{-5cm}\int_{{\tilde{\sigma}}}^{{\tilde{\sigma}}_f} {d\tilde{\sigma}}{|{\tilde{\sigma}|}^{-p}}=
\end{equation*}
\begin{equation}
\left\{ \begin{aligned}
&\frac{1}{-p+1}({|{{\tilde{\sigma}}_f}|}^{-p+1}-{|{\tilde{\sigma}}|}^{-p+1}) ~~\,~\text{for}~~p\ne 1 \\
&~~~~~~~~~~~~~~~~~~~~~~~~~~~~~~~~~~~~~~~~~~~~~~~~~~~~~\,,
\\                                                                    
& \ln(\left|\frac{\tilde{\sigma}_f}{\tilde{\sigma}}\right|) ~~~~~~~~~~~~~~~~~~~\qquad\,\,\quad\text{for}~~p=1 \\
\end{aligned} \right.
\end{equation}

\noindent and

\begin{align}\notag
	&\int_{\tilde{\sigma}}^{{\tilde{\sigma}}_f}d \tilde{\sigma}{{|{\tilde{\sigma}}|}^{-p}}e^{{\tilde{\sigma}}}=
	\\ & ({|{{\tilde{\sigma}}_f}|}^{-p}e^{{\tilde{\sigma}}_f}-{|{\tilde{\sigma}}|}^{-p}e^{\tilde{\sigma}})+
	p\int_{\tilde{\sigma}}^{{\tilde{\sigma}}_f}d\tilde{\sigma}
	{{|{\tilde{\sigma}}|}^{-p-1}}e^{\tilde{\sigma}},
\end{align}

\noindent where $A'\equiv\gamma^{p-2}{A}$. Now, combining Eqs.~(54),~(55) with~(69) and~(70),
the scalar spectral index $n_s$, and tensor-to-scalar ratio $r$
are calculated as

\begin{equation}
n_s=1+\frac{{|{\tilde{\sigma}}|}^{p}}{A'}\big[\frac{-1-p{{\tilde{\sigma}}^{-1}}+c(2-p{{\tilde{\sigma}}^{-1}})e^{\tilde{\sigma}}}
{(1+ce^{\tilde{\sigma}})^2}\big],
\end{equation}

\noindent and

\begin{equation}
r=8\frac{{|{\tilde{\sigma}}|}^{p}}{A'}\big(\frac{1}{1+ce^{\tilde{\sigma}}}\big)^2,
\end{equation}

\noindent respectively. According to our investigations, inflation occurs in the branches $\tilde{\sigma}>0$ ($\sigma$ and $\gamma$ have same signs) and $\tilde{\sigma}<0$ ($\sigma$ and $\gamma$ have opposite signs) for $p>0$ and $p<0$, respectively. Note that $\sigma_f$ and $\sigma$ are obtained inversely proportional to $\gamma$, for any specific $c$ and $p$, implying that $r(n_s)$ and ${N_e}(n_s)$ are independent of $\gamma$.  

In Fig.~(\ref{fig2}), $r(n_s)$ and ${N_e}(n_s)$ are shown for $V_0 e^{-\tilde{\sigma}}$, $p>0$, and two values of $c$. Depending on the values of $p$ and $c$, the curves of $r(n_s)$ lie within the 68\% or 95\% confidence regions of the Planck 2018 TT,TE,EE+lowE+lensing+BK18+BAO data. When $p>8$ and $c=0$,~$r(n_s)$ increases, but $c\not=0$ leads to $\tilde{\sigma}_f=\frac{p}{2}$ and ${A}=0$, undifinable expressions for $\varphi$ (56) and canonicalized potential~(68), and zero for the kinetic term in the Lagrangian~(1). Furthermore,~$r(n_s)$ approaches zero when $p$ tends to zero. Similarly, $r(n_s)$ and ${N_e}(n_s)$ have been plotted in Fig.~(\ref{fig3}) for $V_0 e^{-\tilde{\sigma}}$, $p<0$, and two values of $c$. The curves of $r(n_s)$ lie within the 68\% or 95\% confidence regions of the Planck 2018 TT,TE,EE+lowE+lensing data, again depending on the values of $p$ and $c$. When $p<-80$,~$r(n_s)$ and ${N_e}(n_s)$ decrease gradually, and as $p$ approaches zero, $r(n_s)$ gets away from observations, and ${N_e}(n_s)$ grows rapidly. As shown in Figs.~(\ref{fig2}) and (\ref{fig3}), for $p>0$,~$r(n_s)$ decreases by increasing in $c$, and thus, the curves of $r(n_s)$ are more consistent with the 68\% confidence region of the Planck 2018 TT,TE,EE+lowE+lensing+BK18+BAO data. For $p<0$,~$r(n_s)$ increases as a function of $c$, and consequently, the consistency between $r(n_s)$ and observations is reduced. It is worth noting that the evaluation $r(n_s)$ versus $c$ becomes more clear as the values of $p$ increase.
\section{Summary and discussions}

dRGT theory cannot provide a unique source for both dark energy
and inflation. Motivated by this shortcoming of dRGT theory, and
also, the successes of Lagrangian~(1) in describing the primary and current inflationary eras of the Universe \cite{25,26,27,28,29,30,31}, we addressed some inflationary scenarios in the
framework of dRGT theory. Throughout our analysis, we only focused
on the power and exponential potentials, both of which can
satisfactorily produce 50-70 number of e-foldings, required to
solve the flatness problem. Also, comparing the curves of $r(n_s)$ with the
Planck 2018 data \cite{34} and its combination with BK18 and BAO \cite{35}, it has been obtained that the established
models provide acceptable outcomes.

\nocite{*}
\bibliography{apssamp}

\end{document}